\documentclass{cimento}

\usepackage{epsfig}
\title{Have the missing cosmic baryons been found?}
\author{E.~Behar,\from{ins:Technion}\ETC,
S.~Dado\from{ins:Technion}, 
A.~Dar\from{ins:Technion}, A.~Laor\from{ins:Technion}}
\instlist{\inst{ins:Technion} Department of Physics,
Technion Israel Institute of Technology, Haifa 32000, 
Israel}

\PACSes{Cosmology,~98.80.-k, Gamma ray bursts,~98.70.Rz,  
Quasars,~98.54.Aj}

\begin{document}

\maketitle

\begin{abstract}

The angular power spectrum of the cosmic microwave background radiation 
(CMB), the relative abundances of primordial hydrogen, deuterium and 
helium isotopes, and the large-scale structure of the universe all 
indicate that 4.5\% of the current mass density of the universe consists 
of baryons. However, only a small fraction of these baryons can be 
accounted for in stars and gas inside galaxies, galaxy groups and galaxy 
clusters, and in spectral-line absorbing gas in the intergalactic medium 
(IGM). Too hot to show up in Lyman-absorption, too cool to cause 
detectable spectral distortions of the cosmic microwave background 
radiation, and too diffused to emit detectable X-rays, about 90\% of the 
cosmic baryons remain missing in the local universe (redshift 
$z\!\sim\!0$). Here, we report on prevalent, isotropic, source 
independent, and fairly uniform soft X-ray absorption along the lines of 
sight to high-z gamma-ray bursts (GRBs) and quasars. It has the magnitude, 
redshift and energy dependence that are expected from a hot diffused IGM 
that contains the missing cosmological baryons and has a mean metallicity 
similar to that in the intracluster medium (ICM) of galaxy clusters.

\end{abstract}

\section{Introduction} 

The intergalactic medium (IGM) is extremely difficult to observe. Its 
tremendously low density and high temperature are believed to elude most 
absorption and emission detection. Thus, the observed extragalactic 
absorption of light from gamma ray bursts (GRBs) and quasars, the most 
luminous transient and persistent sources in the universe and the farthest 
observable objects in it, may be the only way to probe the IGM. The
observed absorption was usually assumed to take place mainly in the 
neutral interstellar medium within their host galaxies (HGs).  But, in 
many cases the equivalent hydrogen column densities that were inferred 
from their measured UVO 
and soft X-ray spectra were very different and uncorrelated. Such 
discrepancies were found both for distant GRBs \cite{ref:Jakobsson2006, 
ref:Campana2006, ref:Watson2007, ref:Campana2010, ref:Rau2010} and distant 
quasars (see, {\it e.g.,} \cite{ref:Fabian2001, ref:Worsley2004a, 
ref:Worsley2004b, ref:Yuan2005} and references therein). In contrast, the 
metal abundances and column densities of intervening absorbers on the 
sightlines to galactic and nearby extragalactic sources inferred from soft 
X-ray and Lyman-$\alpha$ absorption, in general do not yield such 
discrepant column densities (see, {\it e.g.,} \cite{ref:Watson2007} and 
references therein).

The extragalactic absorption of soft X-rays from GRBs and quasars at small 
redshifts is usually dominated by absorption in their host galaxy. 
However, at large $z$, the soft X-ray opacity of absorbers 
are expected to decrease 
rapidly with $z$ because both the mean metallicity and the photoabsorption 
cross section at an observed energy decrease rapidly with $z$. In 
contrast, the mean opacity of the IGM to soft X-rays is dominated by 
absorption at small redshifts. It increases rapidly with increasing $z$ to 
its asymptotic value $\tau(E)$ independent of $z$ beyond $z\!\sim\!2$. 
Hence, it is not correlated to the UV absorption in the host galaxy and 
yields a discrepant column density if assumed to take place in/near the 
host\footnote{At small redshifts the column densities in the host galaxy 
of GRBs or quasars that are inferred from UVO and soft X-ray absorption 
can also differ significantly for a different reason: The 
ionization of electrons in the external atomic shells by the UVO emission 
of GRBs, and of blazars in particular, extends to much larger galactic 
distances than the ionization of the inner shells in the metals 
responsible for the soft X-ray absorption (see, {\it e.g.,} 
\cite{ref:Fabian2001, ref:Campana2006, ref:Watson2007} and references 
therein).}. It was suggested that the discrepant column densities 
resulted
either from misinterpreting flattening of the intrinsic spectral 
distribution of the soft X-rays at low energy as X-ray absorption, or from
the high level of ionization of hydrogen in the absorber in the HG 
(see, {\it e.g.,} \cite{ref:Fabian2001, ref:Worsley2004a, 
ref:Worsley2004b, 
ref:Yuan2005} and references therein). Both interpretations, however, 
required fine tuning in order to reproduce both the $E$ and $z$ dependence 
of the observed low-energy opacity.

In this paper we propose a different origin for the discrepant column 
densities inferred from UVO and X-ray observations of high $z$ GRBs and 
quasars. While the UVO absorption takes place mainly in the neutral gas in 
the host galaxies, we suggest that the absorption of their X-rays takes 
place mainly in the hot intergalactic medium (IGM) that contains the bulk 
of the cosmic baryons implied by big bang nucleosynthesis and the observed 
angular power spectrum of the cosmic microwave background and radiation, 
in matter that has the same metallicity as that in the intracluster medium 
(ICM) of galaxy clusters. We show that the opacity of such an IGM can 
explain on average the measured soft X-ray absorption of high $z$ GRBs and 
quasars. It is isotropic, practically independent of source and saturates 
at high $z$, uncorrelated to the UVO absorption in the host and, within 
observational errors, has the magnitude and energy dependence expected for 
the hot IGM of standard cosmology.

\section{Intrinsic host column densities from soft X-ray absorption}

The extragalactic opacities to soft X-rays emitted by GRBs and quasars 
that were measured with the X-ray telescope aboard the Swift satellite 
and with the ROSAT, ASCA, BeppoSAX, Chandra and XMM-Newton satellite, 
respectively, were assumed to be entirely due to the absorption within 
the host galaxies at redshift $z$ although the current X-ray spectra 
contain no redshift information. These opacities were converted to 
equivalent hydrogen column densities ${\rm N_{h,HG}(z)}$ of the GRBs' host 
galaxies along the GRBs sightline, using 
\begin{equation} 
\tau(E,z)=\sigma([1\!+\!z]E)\,N_{h,HG}(z)\, (Z/Z_\odot), 
\label{tauHG} 
\end{equation} 
where $\sigma([1\!+\!z]E)$ is the absorption cross section 
of soft X-rays with energy $[1\!+\!z]E$ per hydrogen atom in the host 
galaxy, assuming a neutral absorber with standard solar  
elemental abundances.
Fig.~1 shows the effective HI column densities of the host galaxies of 
GRBs
with known redshift as measured with the Swift X-ray telescope 
\cite{ref:Evans2009, ref:Campana2010}, assuming the standard   
photospheric solar 
abundances compiled in \cite{ref:Anders1989} and those of radio loud 
quasars as measured with the  X-ray telescopes aboard ASCA 
\cite{ref:Reeves2000}, Chandra \cite{ref:Bassett2004}
and XMM-Newton \cite{ref:Page2005, ref:Yuan2005, ref:Grupe2006} satellites. 
The observed increase of the mean $N_{h,HG}$ with $z$ like $(1\!+\!z)^{2.4}$ 
is in stark contrast with its expected decrease with redshift due to 
the general decline of the mean metallicity with redshift in 
standard galaxy formation and stellar evolution theories
and observed in Lyman-$\alpha$ and damped Lyman-$\alpha$ absorbers 
(see, {\it e.g.,} \cite{ref:Savaglio2009, ref:Kaplan2010} and references 
therein).
Moreover, the photoabsorption cross section above the oxygen K edge at 
$E$=0.54 keV for a neutral absorber with a solar metallicity 
is well described by $\sigma([1\!+\!z]E)\!\approx\! 
\sigma(E)\,(1\!+\!z)^{-2.4}$. Hence, the universal increase of $N_{h,HG}$ 
with $z$ like $(1\!+\!z)^{2.4}$ at large $z$ in both GRB and quasar hosts 
simply reflects the fact that the observed extragalactic opacity for $z>2$ 
tends to an asymptotic value independent of $z$  for GRBs and quasars 
as shown in Fig.~2 and Fig.~3. 
In order to produce the observed $z$-independent opacity  at 
large $z$, either the 
metal column density of HGs of GRBs and quasars by some coincidence  
satisfies 
$N_{h,HG}(z)\, (Z/Z_\odot) \!\propto\!1/\sigma([1\!+\!z]E)\!\propto\! 
(1\!+\!z)^{2.4}$, 
or there is a simpler reason why the extragalactic opacity to 
soft X-rays along the line of sight to GRBs and quasars
becomes independent of z at $z\!>\!2$ and of the X-ray 
source\footnote{We have not included radio quiet quasars in our analysis
because their soft X-ray excess  masks their soft X-ray 
absorption.}.
 
\section{The soft X-ray opacity of the IGM}

A natural origin of a universal, isotropic, and z independent
X-ray opacity that is observed in high $z$ GRBs and quasars is the 
intergalactic medium (IGM) of the standard cosmology that contains the 
bulk of the missing baryons\footnote{The baryon mass fraction in the 
universe that was 
inferred from Big Bang Nucleosynthesis \cite{ref:Steigman2007} and from 
the 
angular spectrum of cosmic microwave background radiation 
\cite{ref:Komatsu2010} is $\Omega_b\!\approx\! 0.045$. Only 10\% of these 
baryons reside in galaxies and galaxy clusters\cite{ref:Fukugita2004}, 
while 
the remaining 90\% 
presumably are still in the IGM in the form of a  hot gas whose 
hydrogen and helium 
are fully ionized.}. Using standard cosmology with  a Hubble constant
$H_0\!=\!71\, {\rm km\, s^{-1}\, Mpc^{-1}}$
and a baryon mass fraction $\Omega_b\!=\!0.045$ 
\cite{ref:Komatsu2010} of which $\!\sim\! 74\%$ are hydrogen nuclei
and only a very small fraction of it ($\!\sim\! 10\%$) resides in  
galaxies and galaxy groups and clusters \cite{ref:Fukugita2004}, the  
mean density of hydrogen 
nuclei in the IGM  is 
$n_h\!\approx 0.67\,\Omega_b\, (3\, H_0^2/8\, \pi\, 
G\, m_p)\, (1\!+\!z)^3,$
and  the opacity of such an IGM to soft X-rays emitted at redshift $z$
with locally observed energy $E$ is given by 
\begin{equation} 
\tau_{IGM}(E,z)\approx 2.21\times 10^{21}\, {\rm cm^{-2}}\,\int_0^z  
{\sigma(E,z')\,  
(Z/Z_\odot)\,(1\!+\!z')^3 dz'\over (1\!+\!z')\, 
\sqrt{(1\!+\!z')^3\,\Omega_M +\Omega_\Lambda}}\,,
\label{tauIGM} 
\end{equation}
where $\Omega_M\!=\!0.27$ and 
$\Omega_\Lambda\!=\!0.73$ \cite{ref:Komatsu2010}.
Eq.~(\ref{tauIGM}) predicts a saturation of $\tau$ for $z>2$ since for 
$E\!<\!10$ keV the photo-absorption cross section $\sigma(E)$ scales 
roughly as $E^{-2.4}$, yielding for a redshifted absorber 
$\sigma(E,z)\!\sim \!\sigma(E)(1\!+\!z)^{-2.4}$, and $d\tau/dz$, which 
decreases with $z$ more rapidly than $(1\!+\!z)^{-1.9}$.
This saturation of  $\tau(E,z)$ at $z\!>\!2$ 
is very different from the increase of $\tau$ with $z$, 
expected and observed in the universe for Compton scattering and line 
absorption. It is, however, in good agreement with the observed 
saturation of the soft X-ray opacity inferred from spectral observations  
of large $z$ GRBs and quasars, as shown in Fig.~2 and Fig.~3.

Moreover, the mean high $z$ opacity $\tau(E)$ calculated for the IGM of 
the standard cosmological model using the best available priors 
agrees 
well with that inferred from the measured spectra of high $z$ GRBs and 
quasars: The observed metallicity in the intracluster medium (ICM) at low 
$z$ is roughly 
$Z/Z_\odot
\!\approx\!(0.54\! \pm\! 0.10)\, 
(1\!+\!z)^{-1.25\!\pm 
\!0.25}$, {\it e.g.,} \cite{ref:Balestra2007} and references therein.
This mean metallicity of the ICM, seems to describe well also  
the mean metallicity in damped Lyman-$\alpha$ (DLA) absorbers at 
$z\!<\!4$ \cite{ref:Savaglio2009, ref:Kaplan2010}, 
and is consistent \cite{ref:Dado2008} with that expected 
from the mean star formation rate in the 
universe as a function of $z$ \cite{ref:Kistler2009}, although the 
spread in metallicity in DLAs \cite{
ref:Prochaska2003, ref:Savaglio2009, ref:Kaplan2010}, galaxies 
and galaxy clusters \cite{ref:Balestra2007} at any given $z$ is quite 
large, 
probably reflecting different star formation histories in 
different galaxies and protogalaxies.
Assuming a mean IGM metallicity  
identical to that of the ICM \cite{ref:Balestra2007} and
adopting the photoabsorption cross sections per ISM hydrogen  
of  \cite{ref:Wilms2000},
after removing the contribution from neutral hydrogen, 
and of helium at $z\!<2$,  which presumably are fully ionized 
in the hot IGM,
the IGM opacity to X-rays from large $z$ GRBs and quasars 
at energy  above the carbon edge ($E\!>\!0.29$ keV) tends to  
\begin{equation}
\tau_{IGM}(E)\approx 0.49\, (0.54\, {\rm keV}/E)^{2.4}-
((0.26\,{\rm keV}/E)^{2.4}-1)\Theta[(0.54\, {\rm keV}/E)-1]\,,
\label{tauIGMasym}
\end{equation} 
where $\Theta(x)\!=\!0$ if $x\!<\!0$ and $\Theta(x)\!=\!1$ if $x\!>\!0$ 
and $E\!=\!0.54$ keV is the oxygen edge. The approach to this asymptotic 
behaviour of the opacity at large $z$ is well approximated by 
$\tau(E,z)\!=\!\tau(E)\,(1\!-\!(1\!+\!z)^{2.15}).$
Below 0.5 keV the IGM opacity becomes 
strongly dependent on the ionization state of helium.
The above estimates are valid for a uniform IGM. However, for a clumpy 
IGM, the observed opacity can deviate significantly from its mean asymptotic 
value.

\section{Comparison between theory and observations} 

The intrinsic opacity of the hosts of GRBs and quasars as given by 
eq.~(\ref{tauHG}) with $N_{Z, HG}(z)\!=\!(Z_{HG}/Z_\odot)\, Nh_{HG}(0))$ 
and $\!<\!Z_{HG}/Z_\odot)\!>\!\sim\!(1\!+\!z)^{-1.25}$, decreases with 
increasing $z$ like $\tau_{HG}(E,z)\!=\!(1\!+\!z)^{-3.65}\, 
\tau_{HG}(E,0)$. Hence, its mean contribution to the extragalactic opacity 
becomes negligible at large $z$. Consequently, the opacity towards high 
$z$ GRBs and quasars is dominated by the IGM opacity, which is isotropic, 
independent of source and redshift and uncorrelated to the UVO absorption 
in the host. This is demonstrated in Fig.~2 where we compare the soft 
X-ray 
attenuation of the hot IGM which follows from eq.~({\ref{tauIGM}) and the 
attenuation inferred from observations of the high redshift blazars PMN 
J0525-3343 at $z$=4.4 and GB B1428+4217 at $z=$ 4.72 
\cite{ref:Worsley2004a, 
ref:Worsley2004b} with XMM-Newton and of GRB 090423 
\cite{ref:Salvaterra2009, 
ref:Evans2009} at a record redshift $z$=8.26 with the Swift XRT. These 
extragalactic opacities were obtained after subtraction of the Galactic 
absorption using the Galactic HI column densities of 
\cite{ref:Kalberla2005} 
and the ISM cross section per HI atom of \cite{ref:Wilms2000} 
\footnote{The 
ISM metallicity adopted by \cite{ref:Wilms2000} agrees well with the 
updated 
solar metallicity compiled in \cite{ref:Asplund2009}, which is smaller by 
a factor $\!\sim\!1.62$ than that compiled in \cite{ref:Anders1989}.}.
The complex low-energy behaviour of the attenuation in the IGM is caused by 
the dependence of the photoabsorption cross sections on the ionization 
state of the most abundant elements in the hot IGM. It has 
behaviour a much different 
than that of the attenuation in the neutral ISM in our 
galaxy and the host galaxy. This is demonstrated in Fig.~2 where we show 
the expected opacity of a hot IGM where He is stripped of its two 
electrons (HeIII) and a warm IGM where He retains one of its two atomic 
electrons (HeII).

In Fig.~3 we compare our estimate of the mean extragalactic opacity to 
soft 
X-rays, $\tau(E,z)\!=\!\tau_{HG}(E,z)\!+\!\tau_{IGM}(E,z),$ as a function 
of $z$ at $E$=0.5 keV as given by Eqs.~(\ref{tauIGM}) and (\ref{tauHG}) and 
the opacity inferred from observations of GRBs and radio loud quasars 
(RLQs) with a good S/N ratio. The contribution from a host galaxy with an 
arbitrarily chosen large column density $N_{h,HG}\!=\! 10^{22}\, {\rm 
cm^{-2}}$ as a function of redshift $z$ is also shown in Fig.~3. The 
observations include all Swift/XRT PC observations of GRBs with known 
redshift when spectral variability is minimal \cite{ref:Evans2009, 
ref:Campana2010}, observations with ASCA \cite{ref:Reeves2000} of 
relatively 
low $z$ RLQs (due to relatively low sensitivity and limited soft X-rays 
bandpass data) and observations of high $z$ RLQs with Chandra 
\cite{ref:Bassett2004} and with XMM-Newton \cite{ref:Worsley2004a, 
ref:Worsley2004b, 
ref:Page2005, ref:Yuan2005, ref:Grupe2006} of half a dozen high $z$ 
quasars with a
relatively good S/N ratio. Only high latitude observations 
($N_{h,Gal}\!<\!10^{21}\, {\rm cm^{-2}}$ where the absorptin is not 
dominated by the Galactic absorption) were included. Fig.~2 and Fig.~3 
clearly 
show the general trend towards an constant opacity at $z\!>\!2$,  
isotropic and indepensent of the X-ray source, consistent with that 
expected for a diffused IGM of standard cosmology.

Figs.~3 also show a large spread in the extragalactic opacity 
measured in 
low redshift GRBs. Such a spread is expected from the variety 
of GRB host galaxies and of the GRB locations, environment and 
sightlines in them. This spread in short hard bursts (SHBs) is also 
shown in Fig.~3. Most of these SHBs have a very small $z$ where the IGM 
opacity is quite small compared to the intrinsic opacity in the host 
galaxy. Consequently, one expects the opacities of far-off-center SHBs to 
be quite small while those of near center SHBs to be much larger and 
similar to 
those of long GRBs whose massive star progenitors are also found mainly 
near the center of the host galaxy. These trends are clearly seen in

Part of the observed spread at all redshifts results from the 
approximate nature of the modelling of the intrinsic spectra of GRBs, 
and from the 
approximate knowledge of the Galactic HI column density and metallicity 
along their sightlines. 
As expected, at large $z$, where the contribution of the HG becomes 
negligible, the spread seems to become smaller and the theory seems to 
describe well the mean value of the observed opacities. A clumpy  
IGM at low redshifts, whence most of the IGM opacity comes, may also 
contribute significantly to the spread.
In order to test this possibility we have plotted  
in Fig.~4 the extragalactic optical depth 
$\tau$  at 0.5 keV to GRBs with known redshift  
calculated from the Swift/XRT PC-mean "intrinsic' $Nh$ values
\cite{ref:Evans2009}, and the
error-weighted $\!<\tau\!>$ values integrated 
over $\Delta z$=1 bins. Fig.~4 ahows the  striking constancy of $N_h$
with $z$ and its tendency towards the value of  $\!\sim\! 0.4$ for 
$z\!>\!2$.

\section{Conclusion} 
The extragalactic opacity to soft X-rays from GRBs 
and quasars at small redshifts is dominated by absorption in their host 
galaxy. However, the extragalactic opacity to soft X-rays from high $z$ 
GRBs and quasars appears to be dominated by absorption in 
the IGM at $z\!\leq\!2$. The 
opacity is isotropic, independent of redshift beyond 
$z\!\approx\!2$ and of 
source, and is not related or correlated to the UV absorption in the 
host 
galaxy. It yields a discrepant column density of the host, if it is 
erroneously assumed to be associated with it. The low energy  
X-ray attenuation in the hot ionized IGM is different from that 
of the mostly neutral ISM in our Galaxy and in the host galaxy of the 
source. In 
particular, it suggests  \cite{ref:Behar2011}
that the IGM of the local universe 
contains most of the currently missing cosmic baryons 
implied by big bang nucleosynthesis 
\cite{ref:Steigman2007}, the observed angular power spectrum of the cosmic 
microwave background (CMB) radiation \cite{ref:Komatsu2010} and the 
Thomson opacity inferred from its polarization \cite{ref:Komatsu2010}, but 
only $\!\sim\!10\%$ are present in the galaxies, galaxy clusters and UVO 
absorbers in the local universe \cite{ref:Fukugita2004}. Soft X-ray 
spectra of very luminous high-z blazars with a large S/N ratio can provide 
more stringent tests of the IGM origin of the extragalactic opacity to 
soft X-rays from high $z$ quasars and GRBs. They may also help determine 
the mean metallicity and the clumpiness of the ionized IGM.

\acknowledgments
The authors thank S. Kaspi and A. Nusser for useful comments
and suggestions.

\newpage
\centering
\begin{figure}
\epsfig{file=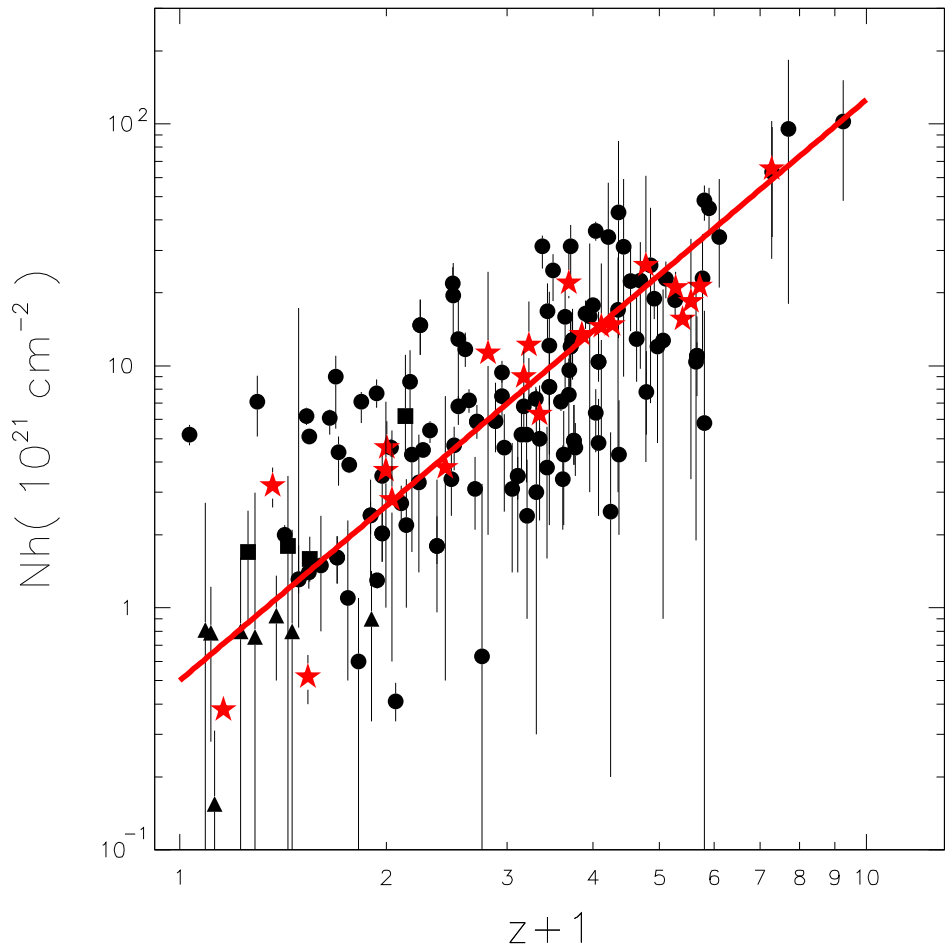,width=14cm,height=14cm}
\caption{Equivalent hydrogen column densities of the HG
of GRBs and radio load quasars as a function of redshift that
were inferred from their absorbed 
soft X-ray spectrum, assuming that the extragalactic absorption took
place in the neutral, solar composition \cite{ref:Anders1989}
ISM of their HG at redshift $z$. 
The GRB data points are from observations  
with the Swift XRT \cite{ref:Campana2010, ref:Evans2009} of long GRBs 
(circles), off-center SHBs (triangles) and near-center SHBs (squares).
The quasar data points (stars) are 
from observations with ASCA \cite{ref:Reeves2000}, Chandra 
\cite{ref:Bassett2004} and XMM-Newton \cite{ref:Worsley2004a, 
ref:Worsley2004b, ref:Page2005, ref:Yuan2005, 
ref:Grupe2006}.}
\end{figure}

\newpage
\begin{figure}
\centering
\epsfig{file=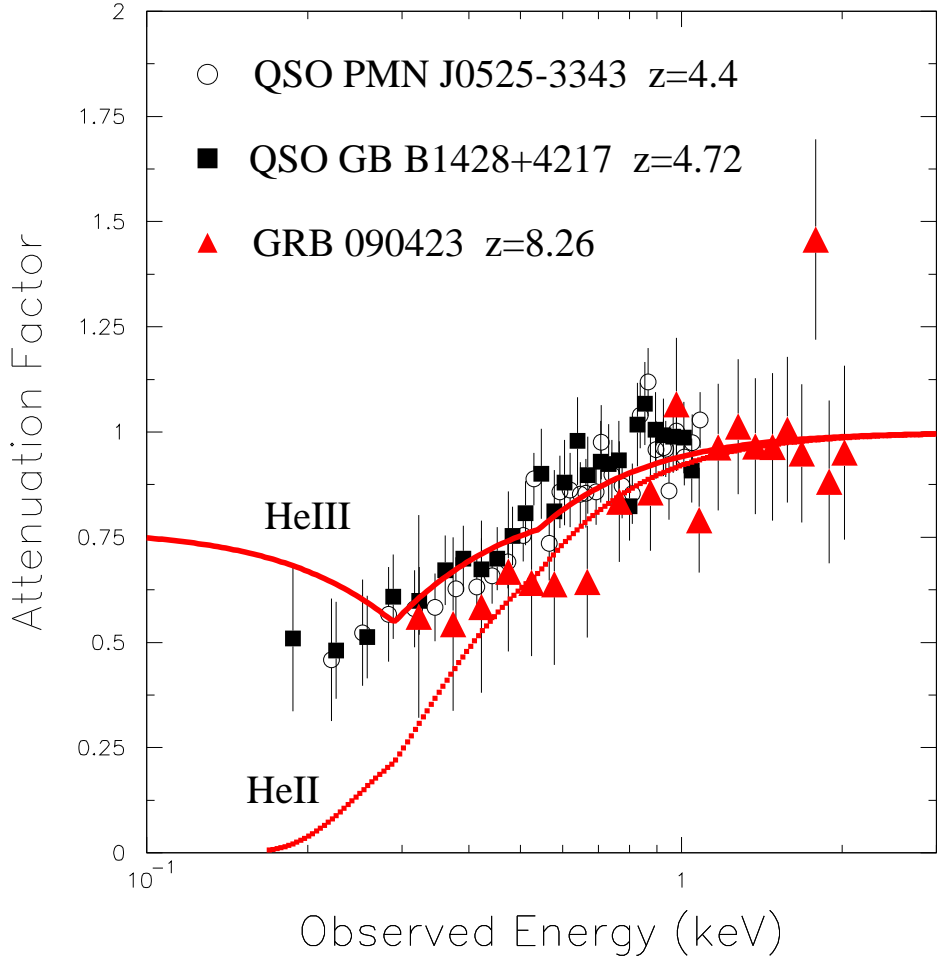,width=14cm,height=14cm}
\caption{Comparison between the extragalactic attenuation of soft X-rays,
$exp(-\tau_{IGM})$,
from the high redshift blazars PMN J0525-3343 at $z$=4.4 
(circles) and
GB B1428+4217 (squares)
at $z=$ 4.72 \cite{ref:Worsley2004a, ref:Worsley2004b}
that was measured 
with XMM-Newton as a function of X-ray energy, 
and that measured with Swift XRT in 
GRB 090423 (triangles) 
at $z$=8.26, the largest measured redshift of a GRB 
\cite{ref:Salvaterra2009, ref:Evans2009}, 
and the attenuation in the hot IGM  of standard cosmology with
the opacity given by  eq.~(\ref{tauIGM}). At energy below 0.5 
keV the IGM opacity 
depends strongly on the ionization state of helium. The upper line (HeIII) 
corresponds to a hot IGM where helium is fully ionized,  
while the lower line (HeII) represents a hot IGM where helium 
is singly ionized. The data show that the absorber is likely in between 
these two cases.} 
\end{figure}

\newpage
\begin{figure}[]
\centering
\epsfig{file=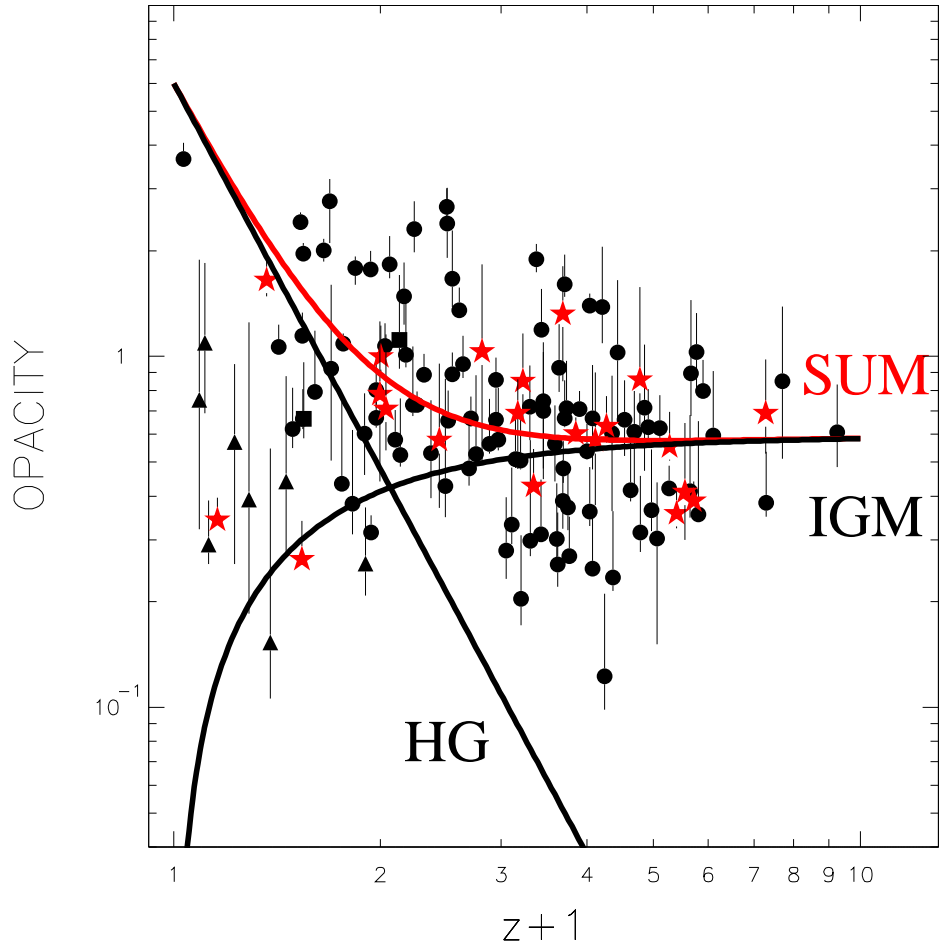,width=14cm,height=14cm}
\caption{Comparison between the extragalactic opacity to soft X-rays
at $E\!\sim\!0.5$ keV as a function of redshift 
measured from GRB and quasar observations  
and the estimated opacity due to
absorption in a hot IGM that contains 90\% of the cosmic baryons with
completely ionized hydrogen and helium    
and partially ionized metals. Circles represent long GRBs,
squares represent near-center SHBs and triangles represent
far-off center SHBs. The GRB data points are from 
afterglow observations with the Swift XRT \cite{ref:Evans2009}
and the quasar data points are from observations  
with ASCA \cite{ref:Reeves2000}, Chandra \cite{ref:Bassett2004}
and XMM-Newton \cite{ref:Worsley2004a, ref:Worsley2004b, ref:Page2005, 
ref:Yuan2005, 
ref:Grupe2006}. The contribution to the extragalactic opacity from a host 
galaxy (HG) with $Nh\!=\!10^{22}\, {\rm 
cm^{-2}}$ at redshift $z$ is also shown.}  
\end{figure}

\newpage
\begin{figure}[]
\hskip -2cm
\epsfig{file=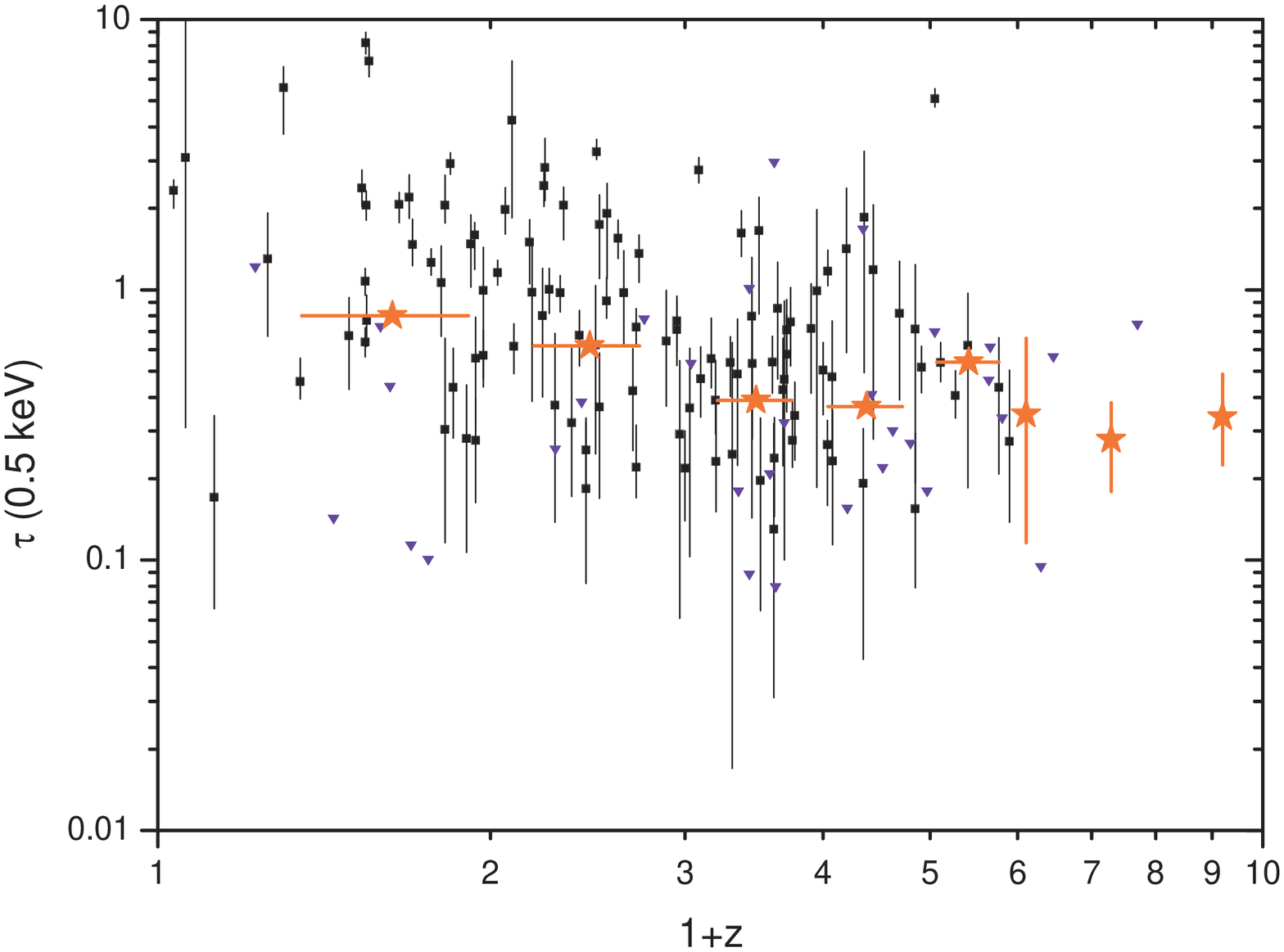,width=14cm,height=18cm,angle=270}
\caption{The extragalactic optical depth
$\tau$  at 0.5 keV to GRBs with known redshift (squares) 
or upper limits (triangles)
calculated \cite{ref:Behar2011} 
from their Swift/XRT PC time-averaged spectrum
\cite{ref:Evans2009}  
and their
error-weighted  average values $\!<\!\tau\!>\!$  within
$\Delta z$=1 bins. Note the  striking constancy of 
$\!<\!\tau\!>\!$ as a function of $z$ for  $z\!>\!2$
and its tendency towards the value $\!\sim\! 0.4$.} 
\end{figure} 

\end{document}